# Intrinsic coherence length anisotropy in nickelate, and some pnictide, and chalcogenide superconductors

Evgeny F. Talantsev[1,2]

[1]M. N. Miheev Institute of Metal Physics, Ural Branch, Russian Academy of Sciences, 18, S. Kovalevskoy St., Ekaterinburg, 620108, Russia

[2]NANOTECH Centre, Ural Federal University, 19 Mira St., Ekaterinburg, 620002, Russia

**Abstract**

Nickelate superconductors, $R_{1-x}A_xNiO_2$ (where R is a rare earth metal and A = Sr, Ca), experimentally discovered in 2019 exhibit many unexplained mysteries as the existence of a superconducting state with $T_c$ (up to 18 K) in thin films and its absence in bulk materials. Another unexplained mystery of nickelates is their temperature-dependent upper critical field, $B_{c2}(T)$, which can be nicely fitted to two-dimensional (2D) models; however the deduced film thickness, $d_{sc,GL}$, exceeds the physical film thickness, $d_{sc}$, by a manifold. To address the latter, it should be noted that 2D models assume that $d_{sc}$ is less than the in-plane and out-of-plane ground state coherence lengths, $d_{sc} < \xi_{ab}(0)$ and $d_{sc} < \xi_c(0)$, respectively, and, in addition, that the inequality $\xi_c(0) < \xi_{ab}(0)$ satisfies. Analysis of the reported experimental $B_{c2}(T)$ data showed that at least one of these conditions does not satisfy for $R_{1-x}A_xNiO_2$ films. This implies that nickelate films are not 2D superconductors, even despite though that the superconducting state is observed only in thin films. Based on this, here we proposed analytical three dimensional (3D) model for global data fit of in-plane and out-of-plane $B_{c2}(T)$ in nickelates. The model is based on a heuristic expression for temperature dependent coherence length anisotropy: $\gamma_\xi(T) = \frac{\gamma_\xi(0)}{1-\frac{1}{a}\times\frac{T}{T_c}}$, where $a > 1$ is a unitless free-fitting parameter. The proposed expression for $\gamma_\xi(T)$, perhaps, has a much broader application because it has been successfully applied to bulk pnictide and chalcogenide superconductors.

# Intrinsic coherence length anisotropy in nickelate, and some pnictide, and chalcogenide superconductors

## I. Introduction

High-temperature superconductivity in Nd$_{1-x}$Sr$_x$NiO$_2$ nickel oxide was experimentally discovered by Li *et al* [1] in 2019, while the first theoretical report in which the oxidation state of Ni$^+$ was established to be a condition under which nickelates become similar to cuprates (and what was made by Li *et al* [1]) was published by Anisimov *et al* [2] in 1999. This experimental discovery initiated further theoretical and experimental studies on R$_{1-x}$A$_x$NiO$_2$ (where R is rare earth, and A = Sr, Ca) thin films [3-50] and bulk [51]. It is widely accepted point of view that the superconducting state in nickelates is exhibited only in thin films [2-51], with a thickness of $d_{sc} \lesssim 15\ nm$. This is one of the primary mysteries in nickelate superconductors.

Another unexplained mystery of nickelates is the temperature dependence of the upper critical field, $B_{c2}(T)$. For instance, when this fundamental field is measured for applied field oriented in perpendicular direction to the (00L) planes of the film (which will be designated as $B_{c2,perp}(T)$, herein), the dependence can be nicely fitted to the Ginzburg-Landau (GL) model [23,36]:

$$B_{c2,perp}(T) = \frac{\phi_0}{2\pi}\frac{1}{\xi_{ab}^2(T)} = \frac{\phi_0}{2\pi}\frac{\left(1-\frac{T}{T_c}\right)}{\xi_{ab}^2(0)}, \tag{1}$$

where $\phi_0$ is superconducting flux quantum, and $\xi_{ab}(0)$ is ground state in-plane coherence length. When external field oriented in parallel direction to the (00L) planes of the film (which will be designated as $B_{c2,para}(T)$), the dependence reported by many research groups can be nicely fitted to the two-dimensional Ginzburg-Landau (2D-GL) model [36,52,53]:

$$B_{c2,para}(T) = \frac{\phi_0}{2\pi}\frac{\sqrt{12}}{d_{sc,GL}}\frac{1}{\xi_{ab}(T)} = \frac{\phi_0}{2\pi}\frac{\sqrt{12}}{d_{sc,GL}}\frac{\sqrt{1-\frac{T}{T_c}}}{\xi_{ab}(0)}, \tag{2}$$

where $d_{sc,GL}$ is the film thickness associated with the use of Eq. 2 for data fit.

This result should be interpreted as direct evidence of 2D superconductivity in nickelates, supporting the experimental observation that the superconducting state is observed only in thin films of nickelates. However, the deduced film thickness $d_{sc,GL}$ (from the $B_{c2,perp}(T)$ and $B_{c2,para}(T)$ data fit to Eqs. 1,2 [23]) by to 2-3 times exceeds the physical film thickness $d_{sc}$ [23]. Here we found that identical problem, i.e. $d_{sc} \not\cong d_{sc,GL}$, does exist for nearly all nickelate films, for which experimental data was reported.

To resolve this issue, here we pointed out that the derivation of Eq. 2 [52] is based on the assumption that physical film thickness $d_{sc}$ is much less than the ground-state coherence length $\xi(0)$ of the superconductor, which means that for anisotropic superconductors:

$$d_{sc} \ll \xi_{ab}(0) \tag{3}$$

and

$$d_{sc} \ll \xi_c(0) \tag{4}$$

where $\xi_c(0)$ is the out-of-plane coherence length. In addition, there is another hidden assumption for the derivation of Eq. 2, which is:

$$\xi_c(0) < \xi_{ab}(0) \tag{5}$$

As shown below, at least one of these conditions (Eq. 3-5) is not satisfied for the studied nickelate films. From this, we concluded that there is an incident, that $B_{c2,para}(T)$ data of nickelate films is nicely approximated by square root of independent variable of two-fluid model:

$$B_{c2,para}(T) \propto \sqrt{1 - \frac{T}{T_c}}, \tag{6}$$

and the deeper physics behind this dependence should be determined.

In this paper, we propose to resolve this problem by accepting the fact that the superconductivity in nickelates is a three-dimensional (3D) phenomenon and, thus, the upper critical field should be described by standard 3D Ginzburg-Landau equations:

$$\begin{cases} B_{c2,perp}(T) = \frac{\phi_0}{2\pi}\frac{1}{\xi_{ab}^2(T)} = \frac{\phi_0}{2\pi}\frac{\left(1-\frac{T}{T_c}\right)}{\xi_{ab}^2(0)} & (7) \\ B_{c2,para}(T) = \frac{\phi_0}{2\pi}\frac{1}{\xi_c(T)}\frac{1}{\xi_{ab}(T)} = \frac{\phi_0}{2\pi}\frac{1}{\frac{\xi_{ab}(T)}{\gamma_\xi(T)}}\frac{1}{\xi_{ab}(T)} = \frac{\phi_0}{2\pi}\gamma_\xi(T)\frac{\left(1-\frac{T}{T_c}\right)}{\xi_{ab}^2(0)} & (8) \end{cases}$$

where $\gamma_\xi(T) = \frac{\xi_{ab}(T)}{\xi_c(T)}$ denotes the temperature-dependent coherence length anisotropy. By experimenting with many analytical functions, we found a remarkably simple and robust heuristic expression for $\gamma_\xi(T)$, which surprisingly enough can also be applied to iron-based superconductors:

$$\gamma_\xi(T) = \frac{\xi_{ab}(T)}{\xi_c(T)} = \frac{\xi_{ab}(0)}{\xi_c(0)}\frac{1}{1-\frac{1}{a}\times\frac{T}{T_c}} = \gamma_\xi(0)\frac{1}{1-\frac{1}{a}\times\frac{T}{T_c}} \quad (9)$$

where $a$ is a free-fitting parameter (varies within a narrow range of $1.2 < a < 2.3$ for all studied superconductors).

**II. The upper critical field definition**

Before Eqs. 7-9 will be applied for $B_{c2}(T)$ data fit, we should clarify the definition of the $B_{c2}(T)$, because different research groups define this fundamental field using different criteria.

In many reports on nickelates, the upper critical field, $B_{c2}(T)$, and, as a direct consequence of it, the coherence length, $\xi(T)$, were defined/deduced from the magnetoresistance curves, $R(T,B)$, by utilizing 50% of normal state resistance criterion, i.e. $\frac{R(T)}{R_{norm}(T)} = 0.5$ (it should be noted that some research groups [49] utilized the criterion of $\frac{R(T)}{R_{norm}(T)} \to 1.0$, which returns the most overestimated $T_c$ and $B_{c2}(T)$ and the most underestimated $\xi_{ab}(0)$ and $\xi_c(0)$ values).

However, in direct experiments performed by Harvey *et al* [35], it was shown that the diamagnetic response in nickelate films is always appeared at temperature well below the zero-resistance temperature, $T_{c,zero}$ (see, for instance, Fig. 2 in Ref. 35). Because diamagnetism is essential and unavoidable property of the superconducting state, the definition of the fundamental superconducting field (i.e., the upper critical field, $B_{c2}(T)$) at the condition at which the superconducting state does not exist (and thus neither the Abrikosov's vortices, nor the phase coherence of the order parameter and the amplitude coherence of the order parameter exist) is incorrect. The definition by $\frac{R(T)}{R_{norm}(T)} = 0.5$, or, by any other similar ratios, except $\frac{R(T)}{R_{norm}(T)} \to 0$, causes many confusions, and the most notable one is the claim that the Pauli limiting field is violated in practically all thin film superconductors [36,54,55]. However, the primary reason for this miserable violation is the definition of the upper critical field, $B_{c2}(T)$, by the criterion at which the superconducting state does not yet exist.

It should be reaffirmed that because the upper critical field, $B_{c2}(T)$, is defined as the magnetic flux density at which the superconducting state collapses and the diamagnetism is essential property of the superconducting state, the definition of the $B_{c2}(T)$ should be made based on the disappearance of diamagnetic response or, if it is impossible to measure, by the $\frac{R(T)}{R_{norm}(T)} \to 0$ criterion. However, it should be mentioned that these definitions have been implemented in very few studies [56-74].

Based on remarkably overestimated $B_{c2}(T)$ values, defined by $\frac{R(T)}{R_{norm}(T)} = 0.5$ or $\frac{R(T)}{R_{norm}(T)} \to 1$ [49,54,55,75-78], and very broad resistive transition width in some thin film superconductors, new effects/phenomena can be claimed (for instance, the Pauli limiting field violation [36,54,55,75-78]). However, these new effects/phenomena can be explained by the misinterpretation of the thermodynamic fluctuations of the phase and the amplitude of the order parameter in superconductors with low charge carrier density [79-81] as the superconducting state. A strongly fluctuating Fermi sea is not an ordered superconducting condensate, where amplitude and phase coherence have been established across the entire sample. Based on this, it is incorrect to apply basic interpretations developed for superconducting condensate (at which non-zero Meissner response should exist or, at least, zero resistance can be measured in experiment) to a system with strong local fluctuations in space and time, which are manifested as several percent drop in the resistance.

In addition to the aforementioned report by Harvey *et al* [35] on $R_{0.8}Sr_{0.2}NiO_2$ (R = La, Pr, Nd), there are several reports performed on other perfect single-phase superconductors, in which the diamagnetic response was detected at $T_{c,dia-onset}$ which is not above (and in many cases well below) the temperature at which the resistance drops to zero, that is $T_{c,dia-onset} \leq T_{c,zero}$ [35,82-90]. These rare, but high quality, experimental reports provide additional evidence for the need to define the $B_{c2}(T)$ by at least the $\frac{R(T)}{R_{norm}(T)} \to 0$ criterion, which is the most accurate experimental value for the superconducting state collapses/emerges if only resistive measurements have been performed for the given sample. This $B_{c2}(T)$ definition has been implemented, but, again, in very rare studies [81,87,91-93] in comparison with majority of studies, where $B_{c2}(T)$ was defined by $0.5 \leq \frac{R(T)}{R_{norm}(T)} < 1$ criterion (see, for instance, Refs. 94,95).

It is interesting to note that Mandal *et al* [84] defined $B_{c2}(T)$ as by $T_{c,dia-onset}$ (and for this definition, the derived $B_{c2}(0) = 3.79\ T$) as well, as by $\frac{R(T)}{R_{norm}(T)} = 0.5$ criterion (and for this definition, the derived $B_{c2}(0) = 5.44\ T$) for bulk $Zr_2Ir$ single crystal. This result demonstrates that $B_{c2}(0)$ defined by $\frac{R(T)}{R_{norm}(T)} = 0.5$ criterion can be overestimated by a factor of 1.4.

However, for Nd$_{0.775}$Sr$_{0.225}$NiO$_2$ nickelate films, this difference is much larger, shown in Fig. 4 in Ref. 23, where the difference between $B_{c2,perp}(T)$ defined by $\frac{R(T=6\,K)}{R_{norm}(T=6K)} = 0.5$ and by $\frac{R(T=6\,K)}{R_{norm}(T=6\,K)} = 0.01$ is approximately five times (the subscripts *perp* and *para* are for the direction of the applied magnetic field to the surface of the films). A much larger difference between $B_{c2}(T)$ defined by different $\frac{R(T)}{R_{norm}(T)}$ criteria was reported by Xiang *et al* [22] for Nd$_{0.8}$Sr$_{0.2}$NiO$_2$ thin films. These observed differences for nickelate films are much larger in comparison with bulk Zr$_2$Ir [87] and bulk FeSe [71] which both exhibit similar $T_c \cong 7\,K$ value and where the difference in $B_{c2}(T)$ defined by different $\frac{R(T)}{R_{norm}(T)}$ criteria is about 40%.

Recently, independent direct confirmation that $T_{c,dia-onset} < T_{c,zero}$ in the nickelate films has been reported by Zeng *et al* [45], who observed the inequality for films with thickness $5.5\,nm \leq d_{sc} \leq 15.2\,nm$ [45].

Based on the above, in this report, we defined the $B_{c2}(T)$ as the lowest possible $\frac{R(T,B)}{R_{norm}(T,B=0)}$ criterion, which can be applied to given experimental $R(T,B)$ datasets (which depend on signal/noise ratio, and other real-world experimental issues).

## III. Results

### 3.1. La$_{0.8}$Ca$_{0.2}$NiO$_2$ film

Chow *et al* [36] reported $B_{c2,perp}(T)$ and $B_{c2,para}(T)$ datasets for La$_{0.8}$Ca$_{0.2}$NiO$_2$ film (which has a physical thickness $d_{sc} = 15\,nm$) defined by $\frac{R(T,B)}{R_{norm}(T,B=0)} = 0.10$, $\frac{R(T,B)}{R_{norm}(T,B=0)} = 0.50$, and $\frac{R(T,B)}{R_{norm}(T,B=0)} = 0.90$ criteria. By following our discussion in the previous section, in Fig. 1 we showed reported $B_{c2,perp}(T)$ and $B_{c2,para}(T)$ datasets defined by $\frac{R(T,B)}{R_{norm}(T,B=0)} = 0.10$ criterion and global data fit to 2D-GL model (Eqs. 1,2). The fit quality is high and the fitting parameters have low mutual dependence. However, deduced film thickness, $d_{sc,GL} = 8.0 \pm 0.1\ nm$, by nearly two times different from physical film thickness $d_{sc} = 15\ nm$, which is a manifestation of general problem associated with utilization of Eqs. 1,2 for nickelate films [23], as we discussed above.

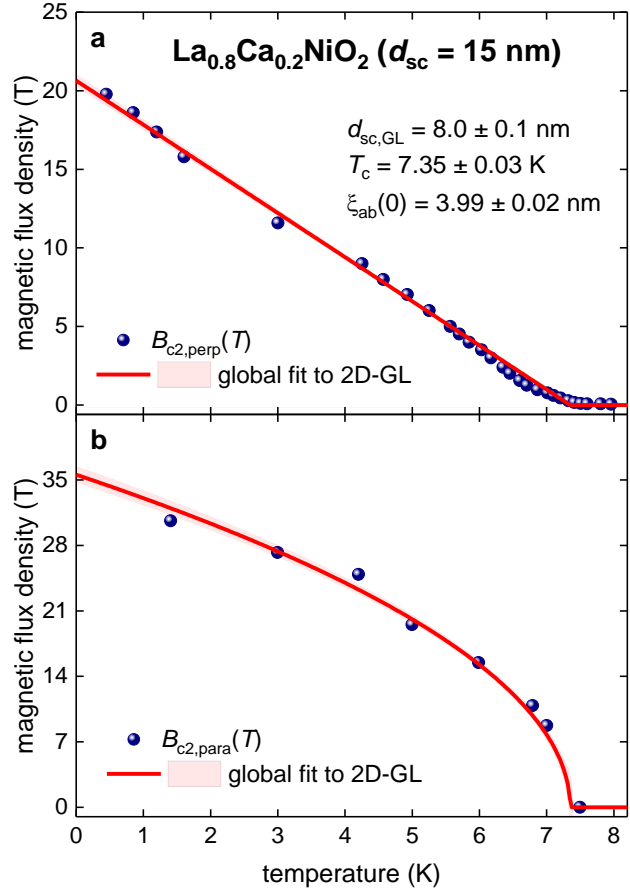

**Figure 1.** (a) $B_{c2,perp}(T)$ and (b) $B_{c2,para}(T)$ data and global fit to the 2D-GL model (Eqs. 1,2) for the La$_{0.8}$Ca$_{0.2}$NiO$_2$ film with a physical thickness $d_{sc} = 15\ nm$. The raw data were reported by Chow et al [36]. Deduced parameters are: $d_{sc,GL} = 8.0 \pm 0.1\ nm$, $T_c = 7.35 \pm 0.03\ K$, $\xi_{ab}(0) = 3.99 \pm 0.02\ nm$. The goodness of fit is (a) 0.9975 and (b) 0.9907. The 95% confidence bands are indicated by pink-shaded areas.

It should be noted that the deduced $\xi_{ab}(0) = 3.99 \pm 0.02\ nm$ is much smaller than any of the two film thicknesses:

$$\xi_{ab}(0) \cong 4\ nm < d_{sc,GL} = 8\ nm \tag{10}$$

$$\xi_{ab}(0) \cong 4\ nm \ll d_{sc} = 15\ nm \tag{11}$$

Considering that there is an expectation that $\xi_c(0) \lesssim \xi_{ab}(0)$, Eqs. 10,11 imply that Eqs. 1,2 cannot be used to fit $B_{c2}(T)$ data for this film, because it is not thin.

In Fig. 2 we show the same reported $B_{c2,perp}(T)$ and $B_{c2,para}(T)$ datasets (as shown in Fig. 1) and the global data fit to our model (Eqs. 7-9).

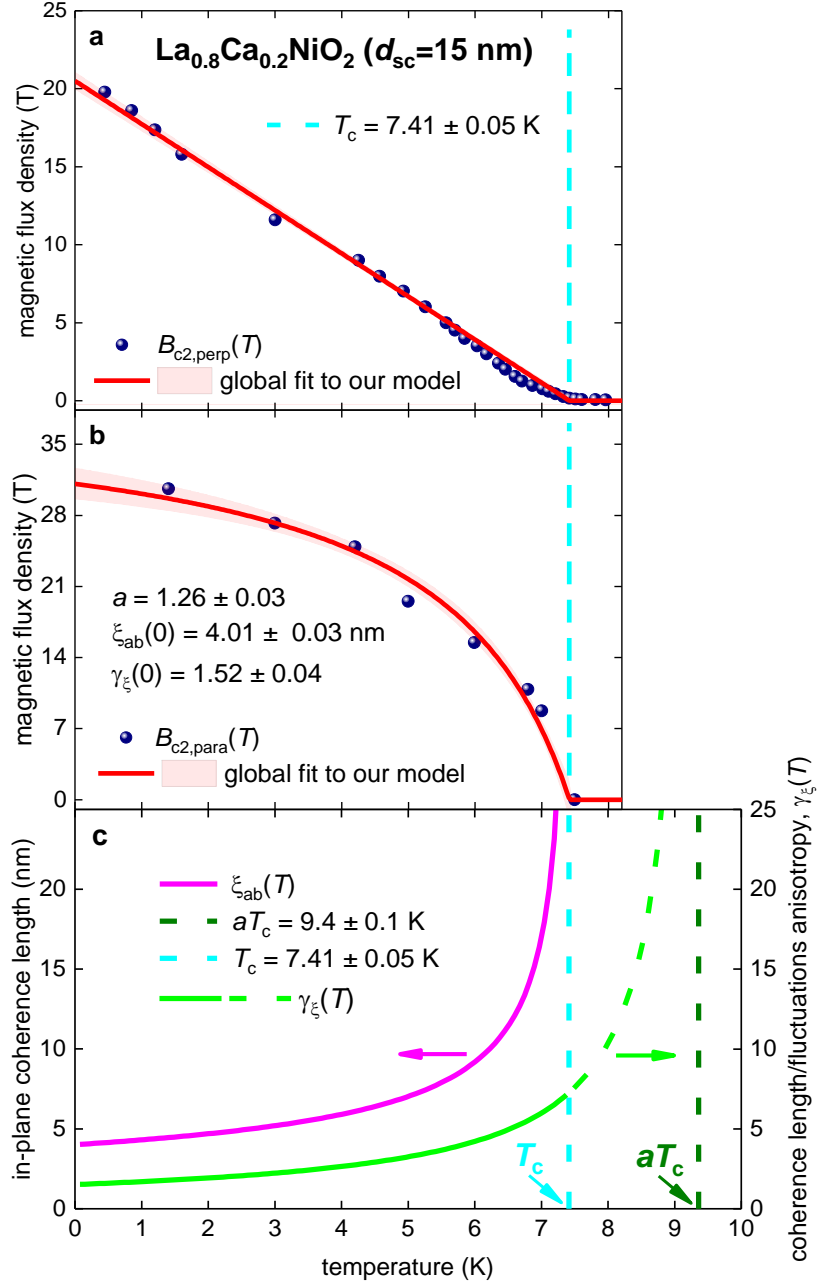

**Figure 2.** Global fit to our model (Eqs. 7-9) for (a) $B_{c2,perp}(T)$, (b) $B_{c2,para}(T)$, and (c) deduced $\xi_{ab}(T)$ and $\gamma_\xi(T)$ for La$_{0.8}$Ca$_{0.2}$NiO$_2$ film with physical thickness $d_{sc} = 15\ nm$. Raw data reported by Chow *et al* [36]. Deduced parameters are: $T_c = 7.41 \pm 0.05\ K$, $\xi_{ab}(0) = 4.01 \pm 0.03\ nm$, $\gamma_\xi(0) = 1.52 \pm 0.04$, $a = 1.26 \pm 0.03$. The goodness of fit is (a) 0.9963 and (b) 0.9832. The 95% confidence bands are indicated by pink shaded areas.

The fit quality is high (Fig. 2) and the mutual dependence of the parameters is low. Deduced coherence lengths are $\xi_{ab}(0) = 4.01 \pm 0.03\ nm$ and $\xi_c(0) = \frac{\xi_{ab}(0)}{\gamma_\xi(0)} = 2.64\ nm$. These values agree with our assumption that the film exhibits three dimensional (3D) superconductivity:

$$\xi_c(0) \cong 2.6\ nm < \xi_{ab}(0) \cong 4\ nm \ll d_{sc} = 15\ nm \tag{12}$$

In Fig. 2(c), we show the calculated temperature-dependent anisotropy of the coherence length, $\gamma_\xi(T) = \gamma_\xi(0) \frac{1}{1-\frac{1}{a}\times\frac{T}{T_c}}$, where all $\gamma_\xi(T)$ values for temperatures in the range of $T_c \leq T < a \times T_c$ are shown by the dotted line.

Physical meaning of this part of the $\gamma_\xi(T)$ curve we discussed in the Discussion section, however in short, it can be mentioned, that the anisotropy should also exist for the phase and the amplitude fluctuations of the order parameter above the transition temperature, $T_c$. Because all nickelates exhibit reasonably wide resistive transitions, similar to other unconventional superconductors (such as cuprates [80] and pnictides [96]), we can propose that $a \times T_c$ value can be interpreted as the onset of the superconducting fluctuations, $T_{fluc}$, in the given material, $T_{fluc} = a \times T_c$ [77-79,96,97].

Thus, our interpretation of the $T_{fluc} = a \times T_c$ is based on the assumption that there is a universal temperature dependence for the anisotropy of the superconducting order parameter and of the fluctuations of this order above the superconducting transition, which, at least from the first glance, looks like a reasonable assumption.

### 3.2. La$_{0.8}$Sr$_{0.2}$NiO$_2$ film

Wang *et al* [98] reported $B_{c2,perp}(T)$ and $B_{c2,para}(T)$ datasets for La$_{0.8}$Sr$_{0.2}$NiO$_2$ film (which has a physical thickness $d_{sc} \sim 7\ nm$) defined by $\frac{R(T,B)}{R_{norm}(T,B=0)} = 0.50$ criterion. In Fig. 3 we show reported $B_{c2,perp}(T)$ and $B_{c2,para}(T)$ datasets and global data fit to the 2D-GL model (Eqs. 1,2). The quality of fit is high, and parameters have low mutual dependence. However, the deduced film thickness, $d_{sc,GL} = 9.5 \pm 0.1\ nm$, exceeds physical film thickness $d_{sc} = 7\ nm$ (Fig. 3), and, in addition, the inequality of:

$$\xi_{ab}(0) \cong 4\ nm < d_{sc} \sim 7\ nm < d_{sc,GL} = 9.5\ nm \qquad (13)$$

shows that the 2D-GL model cannot be used for the analysis (because the film is not sufficiently thin), despite a good fit quality.

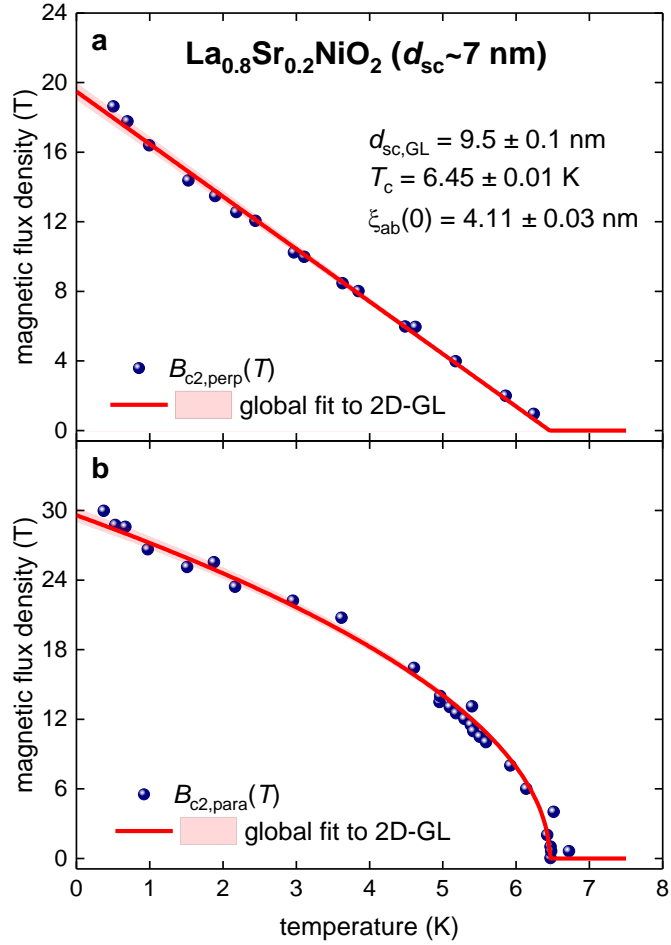

**Figure 3.** (a) $B_{c2,perp}(T)$ and (b) $B_{c2,para}(T)$ data and global fit to the 2D-GL model (Eqs. 1,2) for the La$_{0.8}$Sr$_{0.2}$NiO$_2$ film with a physical thickness $d_{sc} \sim 7\ nm$. Raw data reported by Wang et al [98]. Deduced parameters are: $d_{sc,GL} = 9.5 \pm 0.1\ nm$, $T_c = 6.45 \pm 0.01\ K$, $\xi_{ab}(0) = 4.11 \pm 0.03\ nm$. The goodness of fit is (a) 0.9964 and (b) 0.9837. The 95% confidence bands are indicated by pink-shaded areas.

In Fig. 4 we show the same $B_{c2,perp}(T)$ and $B_{c2,para}(T)$ datasets (as shown in Fig. 3), which were fitted to our model (Eqs. 7-9). The fits quality is high (Fig. 4), and the mutual dependence of the parameters was low. Deduced coherence lengths are $\xi_{ab}(0) = 4.13 \pm$ $0.03\ nm$ and $\xi_c(0) = \frac{\xi_{ab}(0)}{\gamma_\xi(0)} = 2.74\ nm$. These values agree with the assumption of our model that the film exhibits 3D superconductivity:

$$\xi_c(0) \cong 2.7\ nm < \xi_{ab}(0) \cong 4.1\ nm < d_{sc} \sim 7\ nm \tag{14}$$

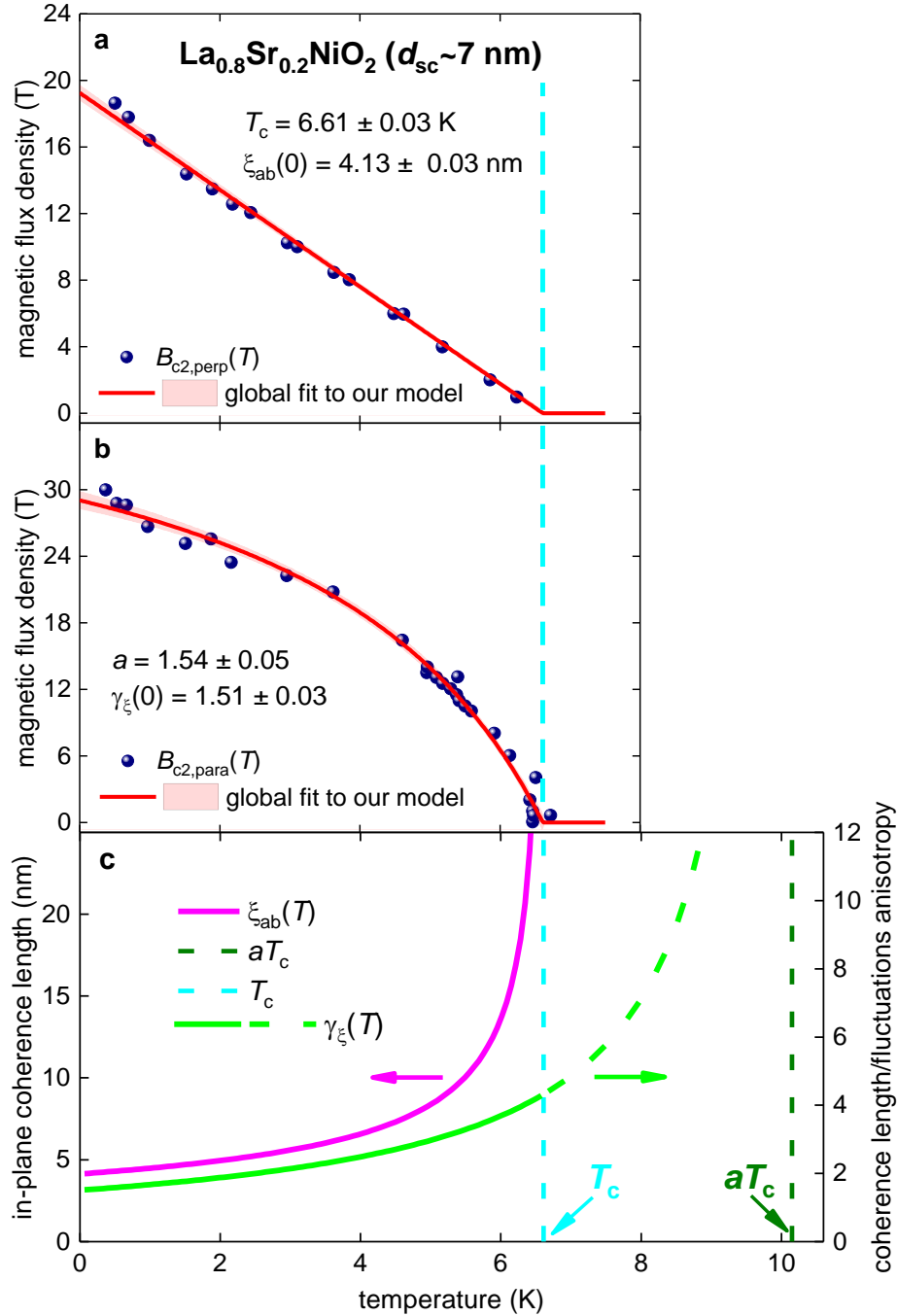

**Figure 4.** Global fit to our model (Eqs. 7-9) for (a) $B_{c2,perp}(T)$, (b) $B_{c2,para}(T)$, and (c) deduced $\xi_{ab}(T)$ and $\gamma_\xi(T)$ for La$_{0.8}$Sr$_{0.2}$NiO$_2$ film with physical thickness $d_{sc} \sim 7\ nm$. Raw data reported by Wang *et al* [98]. Deduced parameters are: $T_c = 6.61 \pm 0.03\ K$, $\xi_{ab}(0) = 4.13 \pm 0.03\ nm$, $\gamma_\xi(0) = 1.51 \pm 0.03$, $a = 1.54 \pm 0.05$. The goodness of fit is (a) 0.9959 and (b) 0.9897. The 95% confidence bands are indicated by pink shaded areas.

In Fig. 4(c), we show the calculated temperature-dependent anisotropy of the coherence length: $\gamma_\xi(T) = \gamma_\xi(0) \frac{1}{1-\frac{1}{a}\times\frac{T}{T_c}}$, where all $\gamma_\xi(T)$ values for temperatures in the range of $T_c \leq T < a \times T_c$ are shown by the dotted line.

### 3.3. Pr$_{0.8}$Sr$_{0.2}$NiO$_2$ film

Wang *et al* [98] reported $B_{c2,perp}(T)$ and $B_{c2,para}(T)$ data for a Pr$_{0.8}$Sr$_{0.2}$NiO$_2$ film (with physical thickness $d_{sc} \sim 7\ nm$) defined by $\frac{R(T,B)}{R_{norm}(T,B=0)} = 0.50$ criterion. In Fig. 5 we show reported $B_{c2,perp}(T)$ and $B_{c2,para}(T)$ data, and global data fit to the 2D-GL model (Eqs. 1,2).

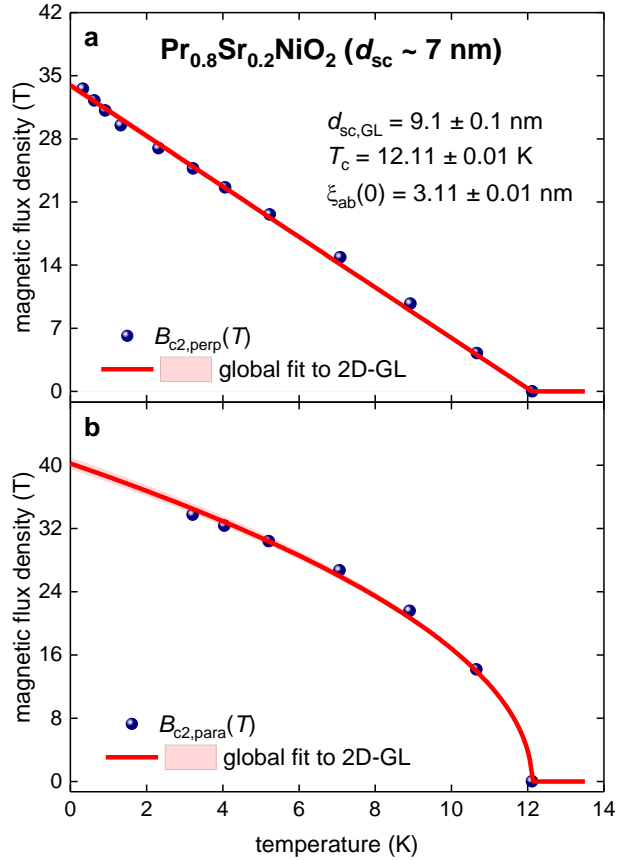

**Figure 5.** (a) $B_{c2,perp}(T)$ and (b) $B_{c2,para}(T)$ data and global fit to the 2D-GL model (Eqs. 1,2) for the Pr$_{0.8}$Sr$_{0.2}$NiO$_2$ film with a physical thickness $d_{sc} \sim 7\ nm$. Raw data reported by Wang *et al* [98]. Deduced parameters are: $d_{sc,GL} = 9.1 \pm 0.1\ nm, T_c = 12.11 \pm 0.01\ K, \xi_{ab}(0) = 3.11 \pm 0.03\ nm$. The goodness of fit is (a) 0.9982 and (b) 0.9976. The 95% confidence bands are indicated by pink shaded areas.

In Fig. 6 we show the same $B_{c2,perp}(T)$ and $B_{c2,para}(T)$ datasets (as shown in Fig. 5), which were fitted to our model (Eqs. 7-9).

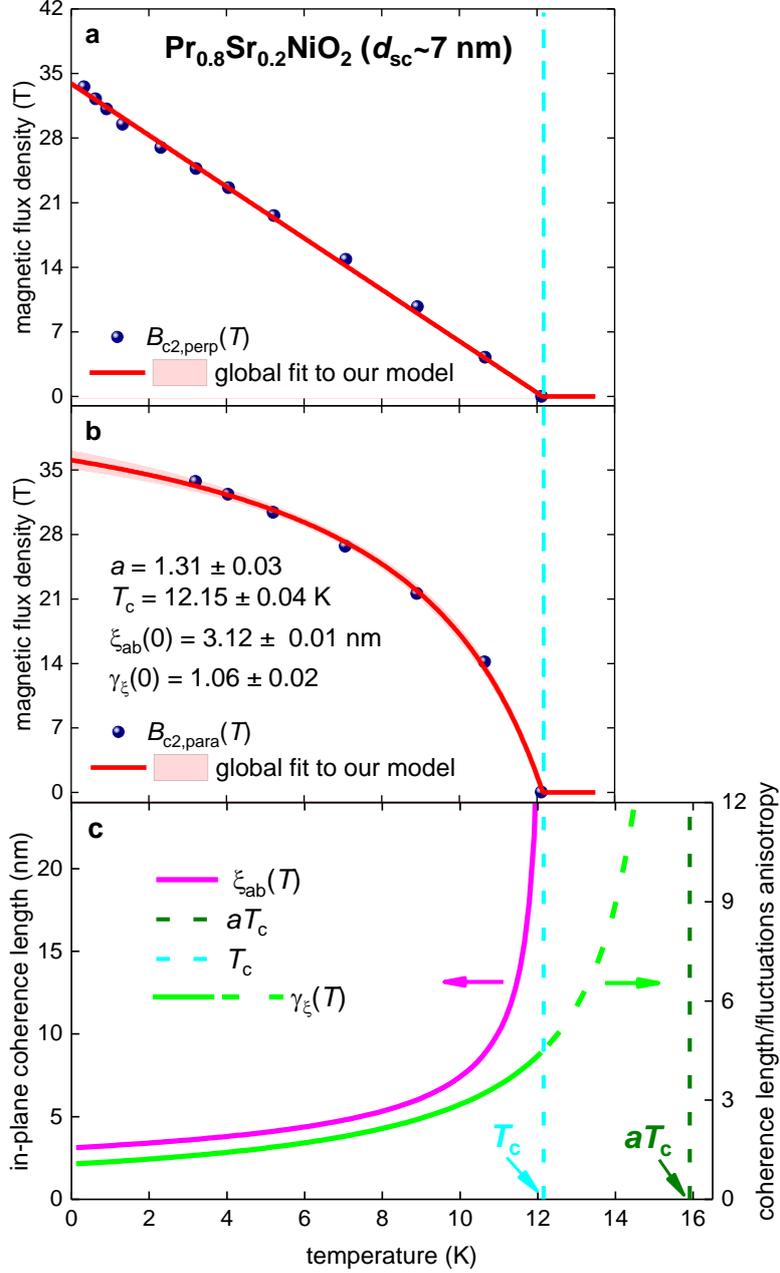

**Figure 6.** Global fit to our model (Eqs. 7-9) for (a) $B_{c2,perp}(T)$, (b) $B_{c2,para}(T)$, and (c) deduced $\xi_{ab}(T)$ and $\gamma_\xi(T)$ for $Pr_{0.8}Sr_{0.2}NiO_2$ film with physical thickness $d_{sc} \sim 7\ nm$. Raw data reported by Wang *et al* [98]. Deduced parameters are: $T_c = 12.15 \pm 0.04\ K$, $\xi_{ab}(0) = 3.12 \pm 0.01\ nm$, $\gamma_\xi(0) = 1.06 \pm 0.02$, $a = 1.31 \pm 0.03$. The goodness of fit is (a) 0.9984 and (b) 0.9985. The 95% confidence bands are indicated by pink shaded areas.

Overall, inequalities (similar to those obtained for the other nickelates (Eqs. 10-14)) were revealed for the $Pr_{0.8}Sr_{0.2}NiO_2$ film:

$$\xi_c(0) \cong 2.9\ nm < \xi_{ab}(0) \cong 3.1\ nm < d_{sc} \sim 7\ nm < d_{sc,GL} \cong 9.1\ nm \qquad (15)$$

### 3.4. Nd$_{0.825}$Sr$_{0.175}$NiO$_2$ film

Wang *et al* [98] reported $B_{c2,perp}(T)$ and $B_{c2,para}(T)$ datasets for Nd$_{0.825}$Sr$_{0.175}$NiO$_2$ film (which has physical thickness $d_{sc} \sim 7\ nm$) defined by $\frac{R(T,B)}{R_{norm}(T,B=0)} = 0.50$ criterion. In Fig. 7 we show reported $B_{c2,perp}(T)$ and $B_{c2,para}(T)$ datasets and global data fit to the 2D-GL model (Eqs. 1,2).

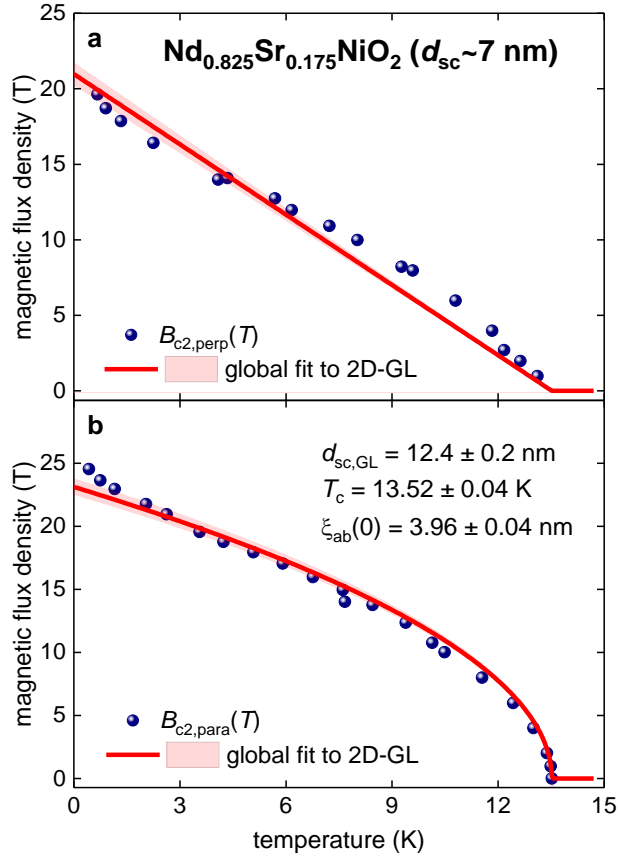

**Figure 7.** (a) $B_{c2,perp}(T)$ and (b) $B_{c2,para}(T)$ data and global fit to the 2D-GL model (Eqs. 1,2) for the Nd$_{0.825}$Sr$_{0.175}$NiO$_2$ film with a physical thickness $d_{sc} \sim 7\ nm$. Raw data reported by Wang *et al* [98]. Deduced parameters are: $d_{sc,GL} = 12.4 \pm 0.2\ nm$, $T_c = 13.52 \pm 0.04\ K$, $\xi_{ab}(0) = 3.96 \pm 0.04\ nm$. The goodness of fit is (a) 0.9658 and (b) 0.9909. The 95% confidence bands are indicated by pink-shaded areas.

In Fig. 8, we show the same $B_{c2,perp}(T)$ and $B_{c2,para}(T)$ datasets (as shown in Fig. 7) which were fitted to our model (Eqs. 7-9). Deduced inequalities:

$$\xi_c(0) \cong 3.4\ nm < \xi_{ab}(0) \cong 4.0\ nm < d_{sc} \sim 7\ nm, \qquad (16)$$

confirmed the 3D superconductivity of the $Nd_{0.825}Sr_{0.175}NiO_2$ film.

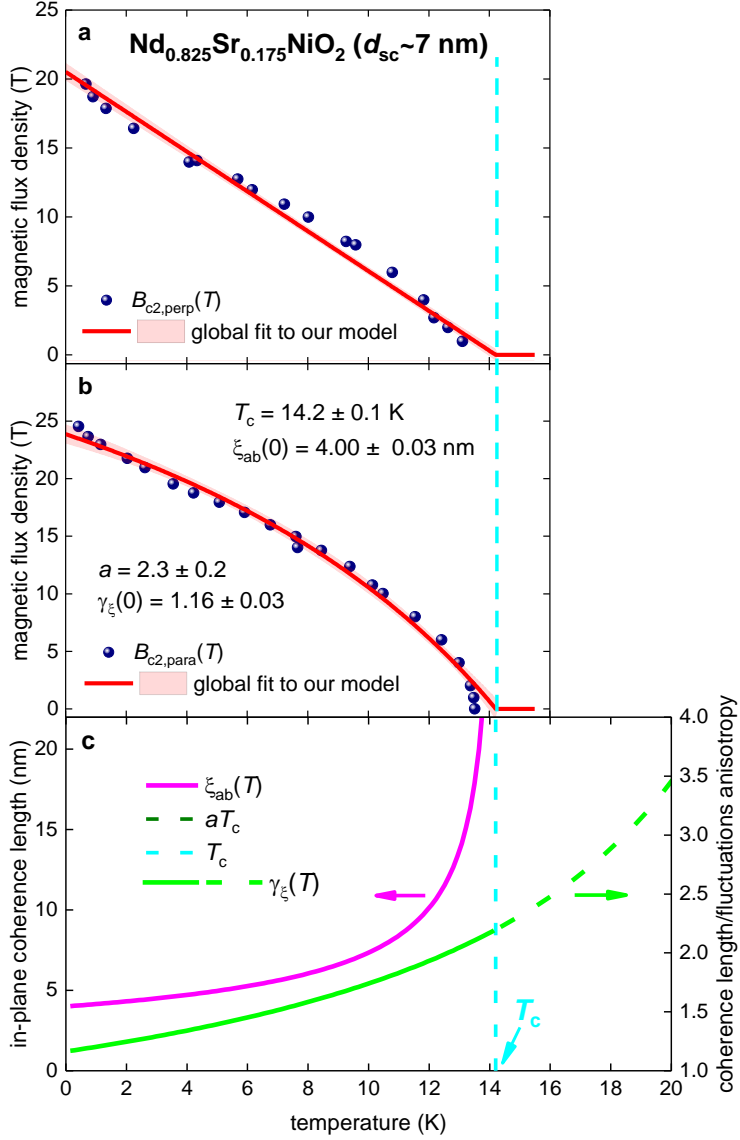

**Figure 8.** Global fit to our model (Eqs. 7-9) for (a) $B_{c2,perp}(T)$, (b) $B_{c2,para}(T)$, and (c) deduced $\xi_{ab}(T)$ and $\gamma_\xi(T)$ for $Nd_{0.825}Sr_{0.175}NiO_2$ film with physical thickness $d_{sc} \sim 7\ nm$. Raw data reported by Wang *et al* [98]. Deduced parameters are: $T_c = 14.2 \pm 0.1\ K$, $\xi_{ab}(0) = 4.00 \pm 0.03\ nm$, $\gamma_\xi(0) = 1.16 \pm 0.03$, $a = 2.3 \pm 0.2$. The goodness of fit is (a) 0.9846 and (b) 0.9900. The 95% confidence bands are indicated by pink shaded areas.

### 3.5. $La_{0.8}Sr_{0.2}NiO_2$ film

Now we return to the $La_{0.8}Sr_{0.2}NiO_2$ compound, for which Wei *et al* [70] recently reported the record high-superconducting transition temperature for nickelates. Wei *et al* [70] also reported $B_{c2,perp}(T)$ and $B_{c2,para}(T)$ datasets defined by $\frac{R(T,B)}{R_{norm}(T,B=0)} = 0.01$, $\frac{R(T,B)}{R_{norm}(T,B=0)} = 0.50$, and $\frac{R(T,B)}{R_{norm}(T,B=0)} = 0.90$ criteria. Despite our understanding that $B_{c2}(T)$ should be defined by the lowest possible $\frac{R(T,B)}{R_{norm}(T,B=0)}$ criterion, in Figs. 9,10 we analysed the $B_{c2}(T)$ data defined by $\frac{R(T,B)}{R_{norm}(T,B=0)} = 0.50$ criterion [70] to make it possible to compare parameters deduced for the La$_{0.8}$Sr$_{0.2}$NiO$_2$ film in Section 3.2 (Figs. 3,4). The film thickness is $d_{sc} \sim 6.8\ nm$ [70], which is practically the same as the one in report by Wang *et al* [98]. In Fig. 9 we show $B_{c2}(T)$ data and global data fit to the 2D-GL model (Eqs. 1,2).

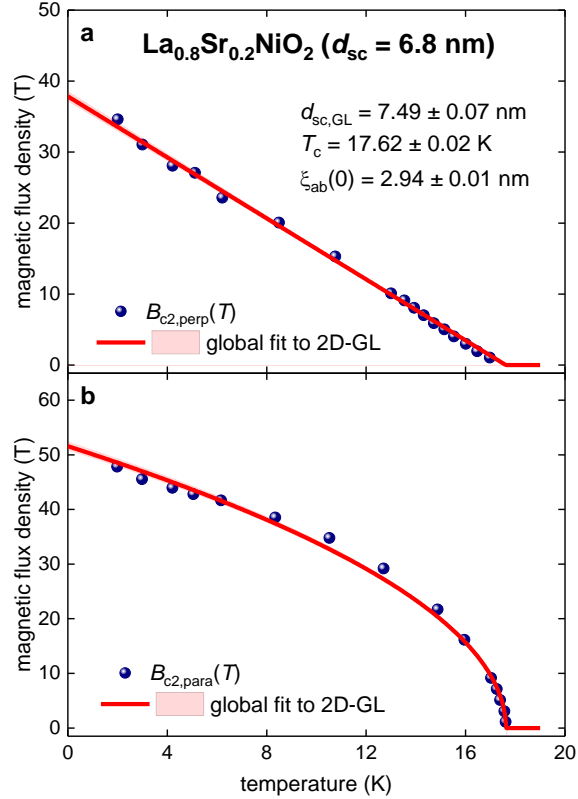

**Figure 9.** (a) $B_{c2,perp}(T)$ and (b) $B_{c2,para}(T)$ data and global fit to the 2D-GL model (Eqs. 1,2) for La$_{0.8}$Sr$_{0.2}$NiO$_2$ film with a physical thickness $d_{sc} = 7\ nm$. Raw data reported by Wei *et al* [70]. Deduced parameters are: $d_{sc,GL} = 7.49 \pm 0.07\ nm$, $T_c = 17.62 \pm 0.02\ K$, $\xi_{ab}(0) = 2.94 \pm 0.01\ nm$. The goodness of fit is (a) 0.9978 and (b) 0.9962. The 95% confidence bands are indicated by pink shaded areas.

In Fig. 10, we show the same $B_{c2,perp}(T)$ and $B_{c2,para}(T)$ datasets (as show in Fig. 9), which were fitted to our model (Eqs. 7-9). Deduced inequalities:

$$\xi_c(0) \cong 2.3\ nm < \xi_{ab}(0) \cong 3.0\ nm < d_{sc} \sim 7\ nm, \qquad (17)$$

confirm the 3D superconductivity of the La$_{0.8}$Sr$_{0.2}$NiO$_2$ film.

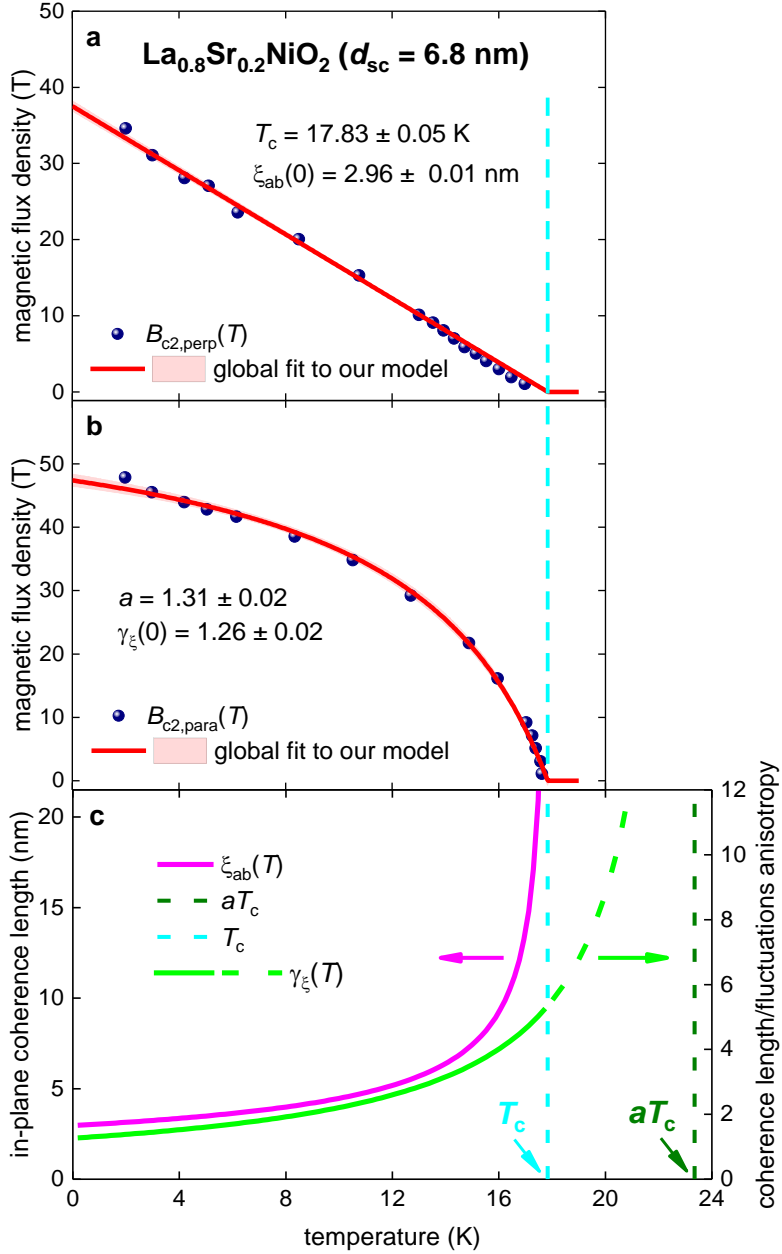

**Figure 10.** Global fit to our model (Eqs. 7-9) for (a) $B_{c2,perp}(T)$, (b) $B_{c2,para}(T)$, and (c) deduced $\xi_{ab}(T)$ and $\gamma_\xi(T)$ for La$_{0.8}$Sr$_{0.2}$NiO$_2$ film with physical thickness $d_{sc} = 6.8\ nm$. Raw data reported by Wei *et al* [70]. Deduced parameters are: $T_c = 17.83 \pm 0.05\ K$, $\xi_{ab}(0) = 2.96 \pm 0.01\ nm$, $\gamma_\xi(0) = 1.26 \pm 0.02$, $a = 1.31 \pm 0.02$. The goodness of fit is (a) 0.9965 and (b) 0.9976. The 95% confidence bands are indicated by pink shaded areas.

In the following sections, we demonstrate that the high quality fit of $B_{c2,perp}(T)$ and $B_{c2,para}(T)$ datasets to Eqs. 1,2 cannot be considered as evidence for 2D superconductivity because we obtained high-quality fits to Eqs. 1,2 for $B_{c2}(T)$ data for bulk iron-based superconductors (IBS).

IBS were experimentally discovered by Hosono's group [99,100] more than 15 years ago, and to the best of our knowledge, there has been no proposal for an analytical expression for the temperature-dependent coherence length anisotropy, $\gamma_\xi(T)$, to this family of superconductors. Here, we show that our 3D model (Eqs. 7-9) can be extended to IBS materials.

### 3.6. Bulk $Tl_{0.58}Rb_{0.42}Fe_{1.72}Se_2$

Jiao *et al* [101] reported $R(T, B_{perp})$ and $R(T, B_{para})$ datasets for bulk single crystal $Tl_{0.58}Rb_{0.42}Fe_{1.72}Se_2$. These authors [101] derived extrapolative values for $B_{c2,perp}(T)$ and $B_{c2,para}(T)$ defined by $\frac{R(T,B)}{R_{norm}(T,B=0)} \to 0.0$ and $\frac{R(T,B)}{R_{norm}(T,B=0)} \to 1.0$ criteria. Because these datasets do not represent values measured in the experiment, in Figs. 11,12 we analysed datasets deduced by $\frac{R(T,B)}{R_{norm}(T,B=0)} = 0.50$ criterion, which represent the measured values.

In Fig. 11, we fitted $B_{c2,perp}(T)$ and $B_{c2,para}(T)$ datasets to Eqs. 1,2 to prove that high-quality fits to the 2D model can be obtained (and, even, "the thickness" of the superconductor, $d_{sc,GL}$, can be deduced) for $B_{c2}(T)$ data measured for bulk anisotropic superconductors.

This implies (Fig. 11) that the thickness, $d_{sc,GL}$, of the "*2D superconductor*" can be deduced from $B_{c2,perp}(T)$ and $B_{c2,para}(T)$ datasets measured for bulk superconductors (see, for instance, Fig. 11) by utilizing the widely used [53] Eqs. 1,2 proposed by Tinkham in the 1960-s [52,102].

Thus, we argue that Eqs. 1,2 are incorrect for use in data analysis because these equations represent reasonably flexible fitting functions, which can be used for smooth data approximation for some superconductors. However, one parameter in these equations, that is $d_{sc,GL}$, which exhibits a unit of length, does not have any physical meaning for bulk superconductors.

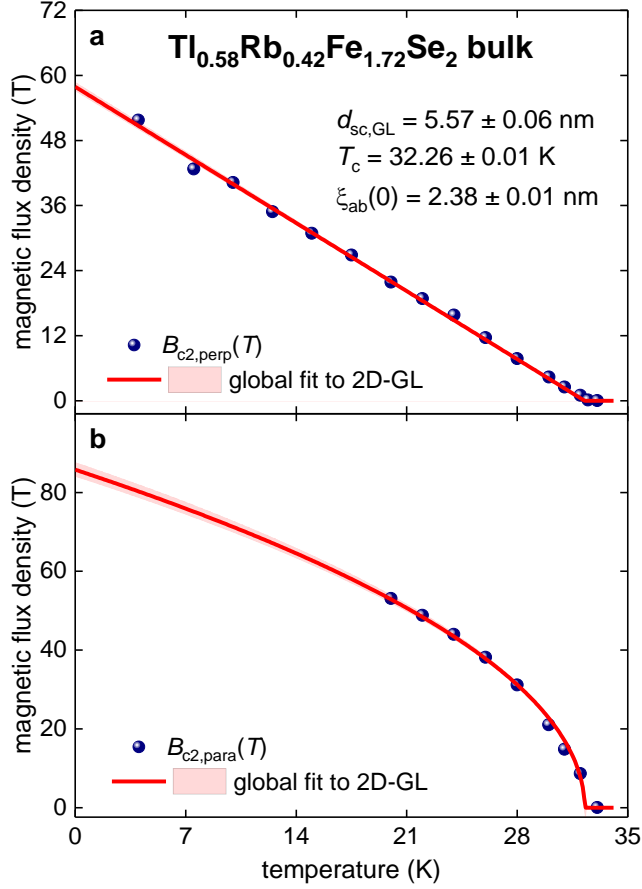

**Figure 11.** (a) $B_{c2,perp}(T)$ and (b) $B_{c2,para}(T)$ data and global fit to the 2D-GL model (Eqs. 1,2) for bulk single crystal $Tl_{0.58}Rb_{0.42}Fe_{1.72}Se_2$. The raw data reported by Jiao *et al* [101]. Deduced parameters are: $d_{sc,GL} = 5.57 \pm 0.06\ nm$, $T_c = 32.26 \pm 0.01\ K$, $\xi_{ab}(0) = 2.38 \pm 0.01\ nm$. The goodness of fit is (a) 0.9986 and (b) 0.9968. The 95% confidence bands are indicated by pink-shaded areas.

This implies that traditional interpretation (see, for instance, Ref. 23) that the violation of the following equation:

$$d_{sc} \cong d_{sc,GL} \qquad (18)$$

in thin film superconductors (including, atomically thin superconductors) should indicate that there is a deep underlying physical effect (for instance, spin-orbit scattering [103,104]), cannot be found to be valid.

In Fig. 12 we show the $B_{c2,perp}(T)$ and $B_{c2,para}(T)$ datasets for $Tl_{0.58}Rb_{0.42}Fe_{1.72}Se_2$ which were fitted to Eqs. 7-9. Fits have high quality. The deduced parameters, in particular:

$$a = 1.29 \pm 0.03, \tag{19}$$

$$\gamma_\xi(0) = 1.24 \pm 0.04, \tag{20}$$

were within the same ranges as those deduced for the nickelate films (Figs. 2,4,6,8,10). This is an evidence that the nickelates exhibit 3D superconductivity.

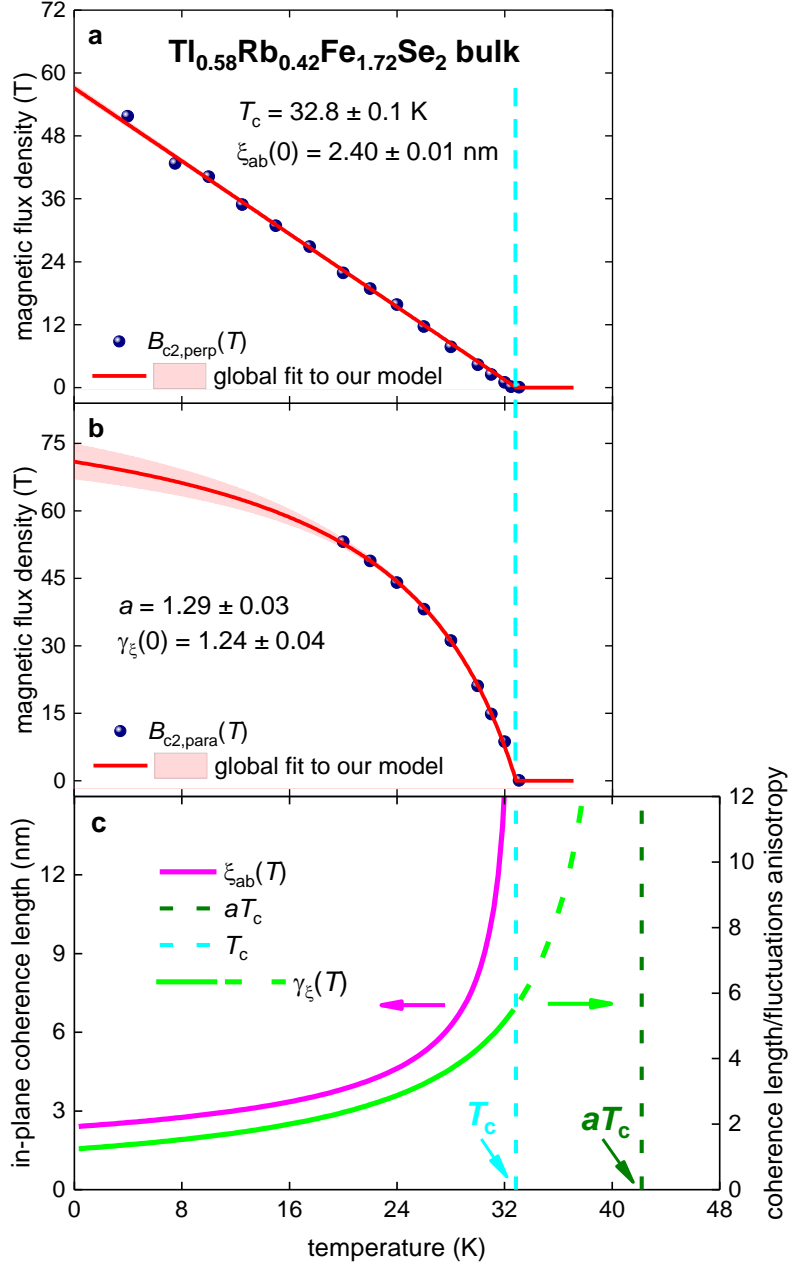

**Figure 12.** The global fit to our model (Eqs. 7-9) for (a) $B_{c2,perp}(T)$, (b) $B_{c2,para}(T)$, and (c) deduced $\xi_{ab}(T)$ and $\gamma_\xi(T)$ for bulk single crystal $Tl_{0.58}Rb_{0.42}Fe_{1.72}Se_2$. The raw data reported by Jiao et al [101]. Deduced parameters are: $T_c = 32.8 \pm 0.1\,K$, $\xi_{ab}(0) = 2.40 \pm 0.01\,nm$, $\gamma_\xi(0) = 1.24 \pm 0.04$, $a = 1.29 \pm 0.03$. The goodness of fit is (a) 0.9984 and (b) 0.9994. The 95% confidence bands are indicated by pink shaded areas.

However, Fig. 12 demonstrates that the 3D model (Eqs. 7-9) can be extended to broader range of superconductors, in particular, on bulk chalcogenides. To demonstrate this, in the next section we apply Eqs. 7-9 on another bulk single crystal chalcogenide superconductor, $Fe_{1.11}Te_{0.6}Se_{0.4}$ [105].

### 3.7. Bulk $Fe_{1.11}Te_{0.6}Se_{0.4}$

Fang et al [105] reported $B_{c2,perp}(T)$ and $B_{c2,para}(T)$ datasets for bulk single crystal $Fe_{1.11}Te_{0.6}Se_{0.4}$ [105] defined by $\frac{R(T,B)}{R_{norm}(T,B=0)} = 0.50$ criterion. In Fig. 13 we fitted $B_{c2,perp}(T)$ and $B_{c2,para}(T)$ datasets to Eqs. 1,2 to demonstrate that high-quality fits to the 2D-GL model can be obtained.

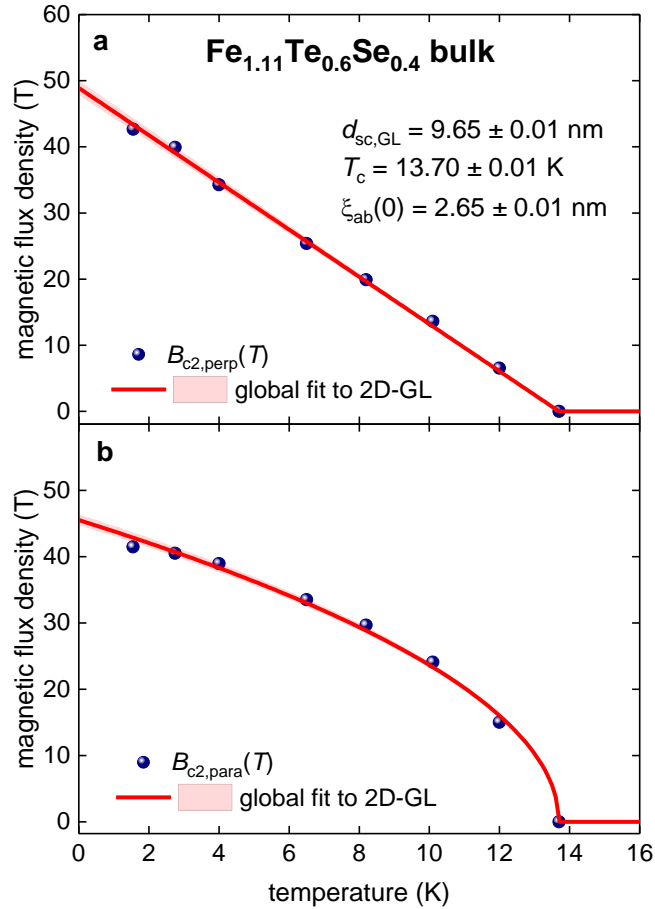

**Figure 13.** (a) $B_{c2,perp}(T)$ and (b) $B_{c2,para}(T)$ data and global fit to the 2D-GL model (Eqs. 1,2) for bulk single crystal $Fe_{1.11}Te_{0.6}Te_{0.4}$. Raw data reported by Fang et al [105]. Deduced parameters are: $d_{sc,GL} = 9.65 \pm 0.01\ nm$, $T_c = 13.70 \pm 0.01\ K$, $\xi_{ab}(0) = 2.65 \pm 0.01\ nm$. The goodness of fit is (a) 0.9986 and (b) 0.9966. The 95% confidence bands are indicated by pink-shaded areas.

Fig. 13 shows that "*the thickness of the 2D superconductor*", $d_{sc,GL}$, of several nanometers (i.e. within a typical range usually deduced for thin film superconductors, including nickeltaes) can be deduced from the fit for this bulk anisotropic superconductor.

In Fig. 14 we show the same $B_{c2,perp}(T)$ and $B_{c2,para}(T)$ datasets (as shown in Fig. 13), which were fitted to Eqs. 7-9.

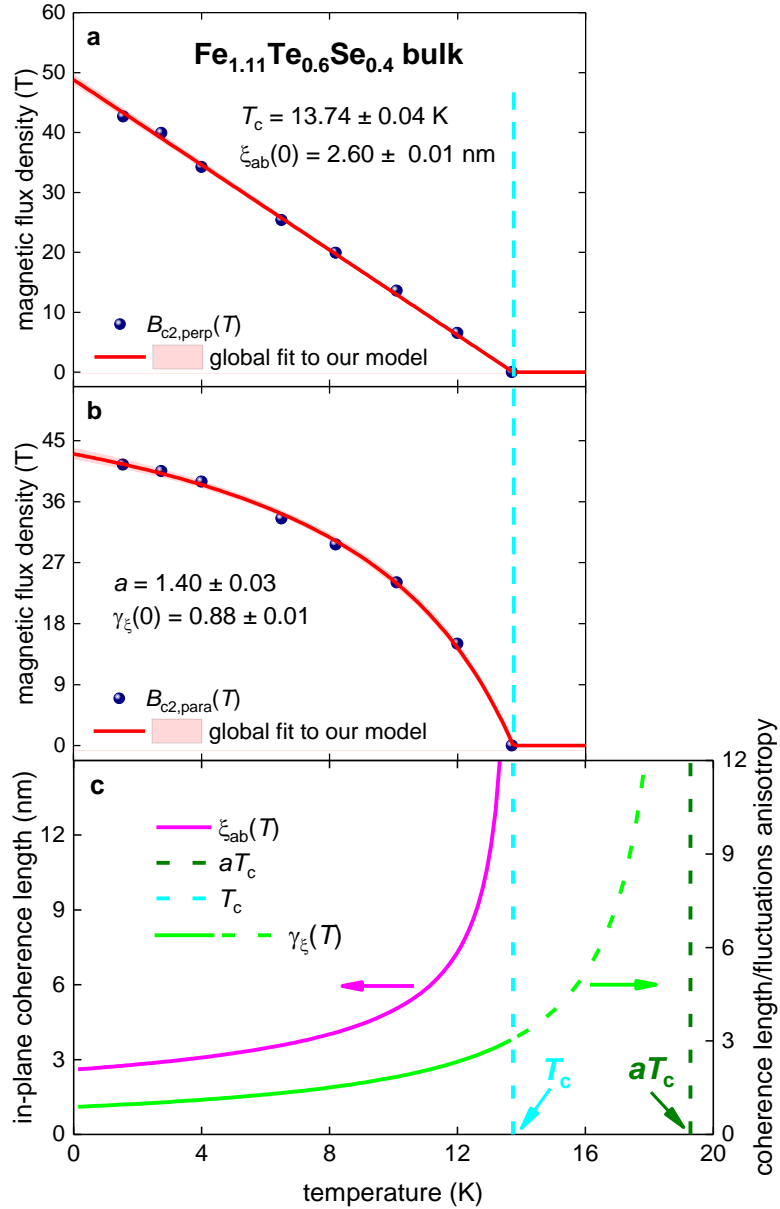

**Figure 14.** Global fit to our model (Eqs. 7-9) for (a) $B_{c2,perp}(T)$, (b) $B_{c2,para}(T)$, and (c) deduced $\xi_{ab}(T)$ and $\gamma_\xi(T)$ for bulk single crystal Fe$_{1.11}$Te$_{0.6}$Se$_{0.4}$. Raw data reported by Fang *et al* [105]. Deduced parameters are: $T_c = 13.74 \pm 0.04\ K$, $\xi_{ab}(0) = 2.60 \pm 0.01\ nm$, $\gamma_\xi(0) = 0.88 \pm 0.01$, $a = 1.40 \pm 0.03$. The goodness of fit is (a) 0.9988 and (b) 0.9989. The 95% confidence bands are indicated by pink shaded areas.

The deduced parameters are in the expected ranges. However, the ground state anisotropy of the coherence length is less than unity:

$$\gamma_\xi(0) = 0.88 \pm 0.01 < 1.0 \qquad (21)$$

It should be noted that $\gamma_\xi(0) < 1.0$ was reported for several iron-based superconductors, and this topic has been discussed (see, for instance Refs. [106,107]).

Owing to the novelty of our model is the Eq. 9, in the following sections we show that the experimental $\gamma_\xi(T) = \frac{B_{c2,para}(T)}{B_{c2,perp}(T)}$ dependences measured for pnictide bulk superconductors can be fitted to Eq. 9, which confirms the validity of the proposed model (Eq. 9).

### 3.8. Bulk KFe$_2$As$_2$

Zocco *et al* [108] reported $B_{c2,perp}(T)$, $B_{c2,para}(T)$, and $\gamma_\xi(T) = \frac{B_{c2,para}(T)}{B_{c2,perp}(T)}$ datasets for bulk single crystal pnictide KFe$_2$As$_2$ superconductor. In Fig. 15, the reported $\gamma_\xi(T)$ dependence is fitted to Eq. 9 (for this fit, we fixed the transition temperature to the value observed in the experiment, $T_c = 3.4\ (fixed)$).

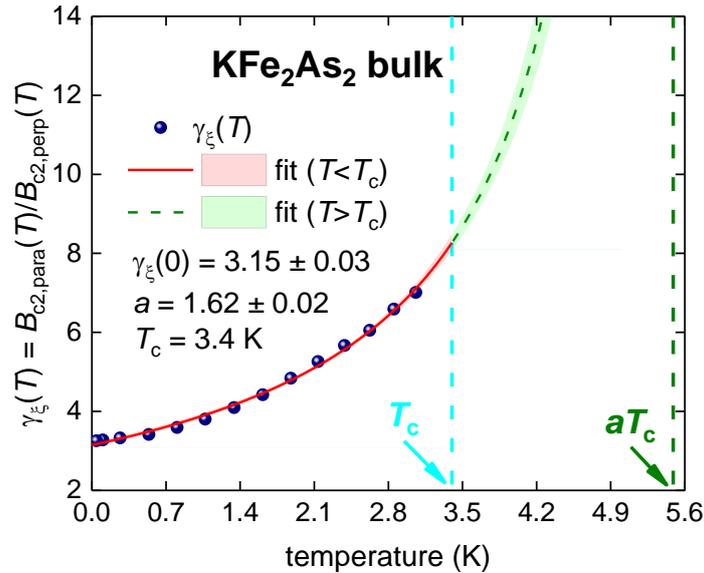

**Figure 15.** $\gamma_\xi(T)$ data and fit to Eq. 9 for single crystal pnictide KFe$_2$As$_2$ superconductor. Raw data reported by Zocco *et al* [108]. Deduced parameters are: $T_c = 3.4\ (fixed)$, $\gamma_\xi(0) = 3.15 \pm 0.03$, $a = 1.62 \pm 0.02$. The goodness of fit is 0.9963. The 95% confidence bands are indicated by pink shaded areas.

The fit has high quality and low parameter dependence. The deduced parameters (Fig. 15) are within the ranges reported above for nickelates and chalcogenides.

### 3.9. Bulk LiFeAs

Khim *et al* [109] and Zhang *et al* [110] reported $B_{c2,perp}(T)$, $B_{c2,para}(T)$, and $\gamma_\xi(T) = \frac{B_{c2,para}(T)}{B_{c2,perp}(T)}$ datasets for bulk single crystal pnictide LiFeAs superconductor. In Fig. 16 we show the fits of the reported $\gamma_\xi(T)$ to Eq. 9 (for this fit, we set the transition temperature to the value observed in the experiments).

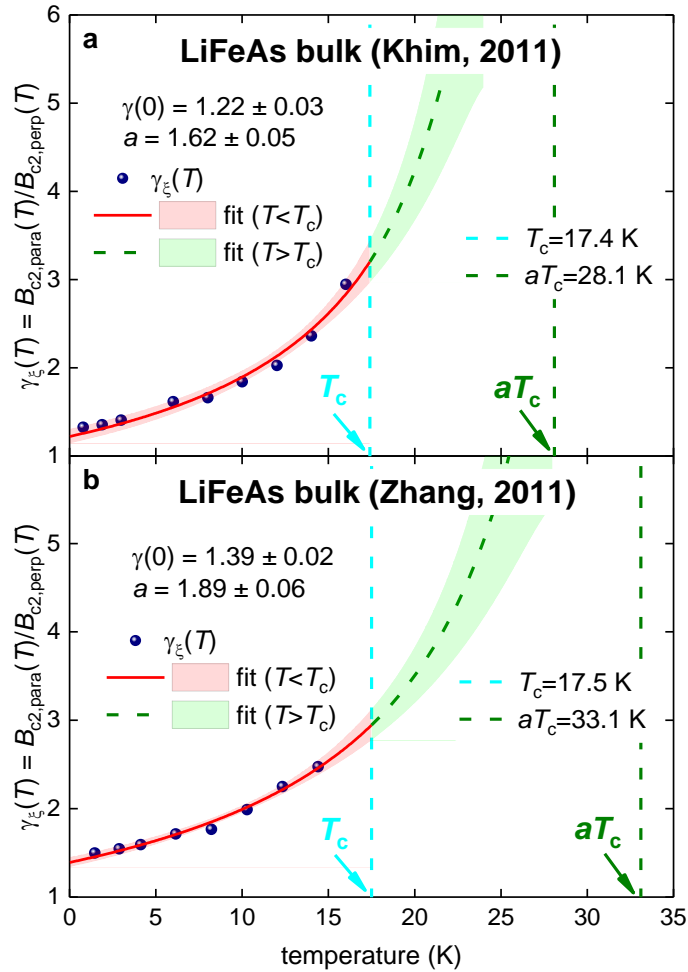

**Figure 16.** $\gamma_\xi(T)$ data and fit to Eq. 9 for single crystal pnictide KFe$_2$As$_2$ superconductor. (**a**) Raw data reported by Khim *et al* [109]. Parameters are: $T_c = 17.4\ (fixed)$ [109], $\gamma_\xi(0) = 1.22 \pm 0.03$, $a = 1.62 \pm 0.05$. The goodness of fit is 0.9777. (**b**) Raw data reported by Zhang *et al* [110]. Parameters are: $T_c = 17.5\ (fixed)$ [110], $\gamma_\xi(0) = 1.39 \pm 0.02$, $a = 1.89 \pm 0.06$. The goodness of fit is 0.9871. The 95% confidence bands are indicated by pink shaded areas.

The fits have a high quality and low parameters dependence. The deduced parameters for two datasets reported by independent research groups are close to each other, within acceptable levels of parameter deviations.

## IV. Discussion

The physical origin of our model (which is primarily based on Eq. 9) can be understood based on an analogy with the temperature-dependent DC magnetic susceptibility, $\chi(T)$, in antiferromagnetic materials [111,112]. The temperature-dependent $\chi(T)$ in any material obeys the Curie-Weiss law (Fig. 17):

$$\chi(T) = \frac{C}{T-\theta} \qquad (22)$$

where $\theta$ is Curie-Weiss temperature, and $C$ is Curie constant.

In the schematic representations of Eq. 21 in Fig. 17, there are three types of magnetic materials that primarily depend on the sign of the Curie-Weiss temperature:

1. $\theta > 0\ K$ for ferromagnetic materials;
2. $\theta = 0\ K$ for paramagnetic materials;
3. $\theta < 0\ K$ for antiferromagnetic materials.

To be consistent with the form of Eq. 21, we can rewrite Eq. 9:

$$\gamma(T) = \frac{aT_c\gamma(0)}{aT_c-T} = -\frac{aT_c\gamma(0)}{T-aT_c} = -\frac{C}{T-\theta} \qquad (23)$$

where $C = aT_c\gamma(0)$, and $\theta = aT_c$.

Despite the negative sign (in K units) of the Curie-Weiss temperature, $\theta$, for antiferromagnetic materials, this value represents one of the fundamental constant of the antiferromagnet, which quantifies the strength of the antiferromagnetic interaction in the material.

In antiferromagnetic materials, the $\chi(T)$ would obey the Curie-Weiss law down to very low temperatures, $T \to 0\ K$ (Fig. 17,c), however, at the Neel temperature, $T_N > 0\ K$, a phase transition occurs, and the $\chi(T)$ does not obey the Curie-Weiss law at $T < T_N$.

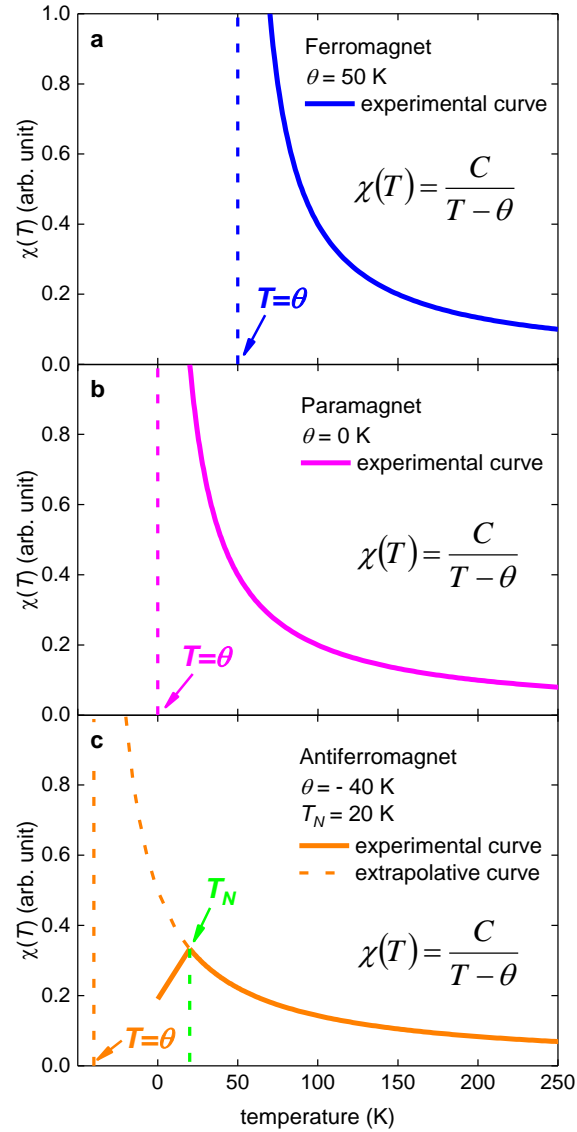

**Figure 17.** Schematic representation of the DC magnetic susceptibility $\chi(T)$ obeys the Curie-Weiss law for (**a**) ferromagnetic, (**b**) paramagnetic, and (**c**) antiferromagnetic materials. $\theta$ is Curie-Weiss temperature (the analogue value in our model is $aT_c$), $C$ is Curie constant (the analogue value in our model is $(aT_c) \times \gamma_\xi(0)$), $T_N$ is Neel temperature (the analogue value in our model is $T_c$).

In our model, $\gamma(T)$ would obey the Eq. 9 up to high temperatures, $T \to aT_c$, however, at $T = T_c$ a superconductor-normal state phase transition occurs, and $\gamma(T)$ becomes undefined at $T > T_c$. However, the latter does not mean that $aT_c$ does not represent any physical value for the material, and our current interpretation of this value is that $T = aT_c$ represents the threshold temperature for the appearance of the anisotropy in the fluctuations of the order parameter in superconductor.

It should also be stressed that our model (Eqs. 7-9) utilizes the simplistic GL expression for the temperature-dependent in-plane coherence length:

$$\xi_{ab}(T) = \frac{\xi_{ab}(0)}{\sqrt{1-\frac{T}{T_c}}}. \tag{24}$$

However, it would be more accurate to use the Werthamer-Helfand-Hohenberg (WHH) theory [113] or its advanced version developed for two-band superconductors by Gurevich [114]. This type of advanced analysis, in conjunction with high-field experimental studies, has been implemented in several studies on IBS [115,116] and nickelates [23].

However, the high flexibility of primary WHH functions (exhibited several parameters, especially for the two-band model), and the nonexistence of WHH functions for $T > T_c$, makes it impossible to extract a simple analytical expression, similar to Eq. 9, for the temperature dependence of the coherence length anisotropy, $\gamma(T)$, which we proposed herein. That is, the fact that:

$$\xi_{ab}(T)|_{T \to T_c} \to \infty, \tag{25}$$

$$\xi_c(T)|_{T \to T_c} \to \infty, \tag{26}$$

does not exclude the following:

$$\gamma_\xi(T) = \frac{\xi_{ab}(T)}{\xi_c(T)}\bigg|_{T \to T_c} \neq \infty. \tag{27}$$

The rigorous mathematical expression for the primary message of this study is as follows:

$$\forall \frac{T}{aT_c} < 1, \exists\, \gamma_\xi(T) \in \mathbb{R}_{>0}, \tag{28}$$

where the standard mathematical symbols are used.

In this regard, the simple heuristic expression for $\gamma_\xi(T)$ (proposed herein (that is Eq. 9)) might be modified to be more accurate (and unfortunately, more complicated); however, our primary message that the anisotropy of the thermodynamic fluctuations in anisotropic superconductors (exhibiting low charge carrier density) is established at some temperature $T = aT_c$ ($a > 1$) above the entire superconducting transition temperature, $T_c$, should remain unchanged.

All superconductors exhibit the second fundamental characteristic length which is the London penetration depth, $\lambda(T)$. In anisotropic superconductors, the $\lambda(T)$ also has two components, which are in-plane London penetration depth, $\lambda_{ab}(T)$, and out of plane London penetration depth, $\lambda_c(T)$. We expect that the same approach, to the described herein for $\gamma_\lambda(T)$, should be applied for the temperature dependent London penetration depth anisotropy, $\gamma_\lambda(T) = \frac{\lambda_{ab}(T)}{\lambda_c(T)}$ [72,117-119]. However, as the model development, as data analysis, as the discussion of this topic are far beyond the frames of the current study.

## V. Conclusion

In this study, we analyzed the temperature dependence of the upper critical field anisotropy and the coherence length anisotropy, $\gamma_\xi(T)$, in nickelate superconductors for which we proposed simple heuristic expression (Eq. 9) within the 3D Ginzburg-Landau model. The proposed expression for $\gamma_\xi(T)$ (Eq. 9) is also applicable to some chalcogenide and pnictide superconductors.

**Funding**

This research was funded by the Ministry of Science and Higher Education of the Russian Federation, grant number No. AAAA-A18-118020190104-3 (theme "Pressure"). The research funding from the Ministry of Science and Higher Education of the Russian Federation (Ural Federal University Program of Development within the Priority-2030 Program) is gratefully acknowledged.